research papers

# In-depth investigations of size and occupancies in cobalt ferrite nanoparticles by joint Rietveld refinements of X-ray and neutron powder diffraction data

Killian Henry,[a,b] Jakob Voldum Ahlburg,[a] Henrik L. Andersen,[a]‡ Cecilia Granados-Miralles,[c] Marian Stingaciu,[a] Matilde Saura-Múzquiz[a]‡ and Mogens Christensen[a]*

[a]Center for Materials Crystallography (CMC), Department of Chemistry and iNANO, Aarhus University, Aarhus, C-8000, Denmark, [b]Université de Lorraine, CNRS, IJL, Nancy, F-54000, France, and [c]Electroceramics Department, Instituto de Ceramica y Vidrio, CSIC, Kelsen 5, Madrid, ES 28049, Spain. *Correspondence e-mail: mch@chem.au.dk

Powder X-ray diffraction (PXRD) and neutron powder diffraction (NPD) have been used to investigate the crystal structure of $CoFe_2O_4$ nanoparticles prepared via different hydrothermal synthesis routes, with particular attention given to accurately determining the spinel inversion degrees. The study is divided into four parts. In the first part, the investigations focus on the influence of using different diffraction pattern combinations (NPD, Cu-source PXRD and Co-source PXRD) for the structural modelling. It is found that combining PXRD data from a Co source with NPD data offers a robust structural model. The second part of the study evaluates the reproducibility of the employed multipattern Rietveld refinement procedure using different data sets collected on the same sample, as well as on equivalently prepared samples. The refinement procedure gives reproducible results and reveals that the synthesis method is likewise reproducible since only minor differences are noted between the samples. The third part focuses on the structural consequences of (i) the employed heating rate (achieved using three different hydrothermal reactor types) and (ii) changing the cobalt salt in the precursors [aqueous salt solutions of $Co(CH_3COOH)_2$, $Co(NO_3)_2$ and $CoCl_2$] in the synthesis. It is found that increasing the heating rate causes a change in the crystal structure (unit cell and crystallite sizes) while the Co/Fe occupancy and magnetic parameters remain similar in all cases. Also, changing the type of cobalt salt does not alter the final crystal/magnetic structure of the $CoFe_2O_4$ nanoparticles. The last part of this study is a consideration of the chemicals and parameters used in the synthesis of the different samples. All the presented samples exhibit a similar crystal and magnetic structure, with only minor deviations. It is also evident that the refinement method used played a key role in the description of the sample.

## 1. Introduction

Permanent magnets (PMs) have a large number of both domestic and industrial applications (Lewis & Jiménez-Villacorta, 2013; Coey, 2002; Cullity & Graham, 2009; Furlani, 2001; Jiles, 2015; López-Ortega, Estrader et al., 2015). They constitute key components in e.g. electric motors/generators (Lewis & Jiménez-Villacorta, 2013; Coey, 2002; Cullity & Graham, 2009; Jiles, 2015), magnetic recording/storage media (Cullity & Graham, 2009; Furlani, 2001; Jiles, 2015) and microphones/loudspeakers (Lewis & Jiménez-Villacorta, 2013; Coey, 2002; Cullity & Graham, 2009; Jiles, 2015; Mathew & Juang, 2007; López-Ortega, Estrader et al., 2015), are used in medical magnetic resonance imaging (Lewis & Jiménez-Villacorta, 2013; Coey, 2002; Cullity & Graham, 2009; Jiles, 2015; Pankhurst et al., 2003), and even have a number of both established and potential biomedical applications (Pankhurst *et al.*, 2003; Valenzuela, 2012; Ramanujan, 2009). The utilization of PMs has expanded drastically over the years, making the PM materials industry a multi-billion-dollar global market (Valenzuela, 2012; Granados-Miralles *et al.*, 2018; Quesada *et al.*, 2016; Coey, 2010*a*). The most powerful PMs are made from materials that contain rare-earth elements (REE), *e.g.* $Nd_2Fe_{14}B$ and $SmCo_5$, but due to political and environmental concerns related to REE mining, there is now an increasing drive and demand for developing REE-free alternatives (Lewis & Jiménez-Villacorta, 2013; López-Ortega, Estrader *et al.*, 2015; Gandha *et al.*, 2015; Granados-Miralles *et al.*, 2016; Abalakin, 2006; European Commission, 2017; Coey, 2011). Even though some REE-containing PMs are irreplaceable due to their unique magnetic properties, REE-free PMs could potentially replace the low-grade REE PMs in some applications (Lewis & Jiménez-Villacorta, 2013; Coey, 2011; Pedrosa *et al.*, 2016; Gutfleisch *et al.*, 2011). In this context, there is a growing interest in PMs based on nanostructured cobalt ferrite, $CoFe_2O_4$ (López-Ortega, Estrader *et al.*, 2015; Leite *et al.*, 2012; Cedeño-Mattei & Perales-Pérez, 2009; Zhao *et al.*, 2008), due to the demonstrated ability to enhance the magnetic properties of the $CoFe_2O_4$ material through nanoparticle size control (Andersen & Christensen, 2015; Stingaciu *et al.*, 2017; Song & Zhang, 2004), crystal engineering (changing the distribution of Co and Fe between the different sites in the spinel structure) (Ahlburg *et al.*, 2020; Yu *et al.*, 2002; Andersen *et al.*, 2019; Na *et al.*, 1993) and partial reduction to form $CoFe/CoFe_2O_4$ exchange-spring nanocomposites (López-Ortega, Estrader *et al.*, 2015; Granados-Miralles *et al.*, 2018; Quesada *et al.*, 2016; Leite *et al.*, 2012; Ahlburg *et al.*, 2020; Kahnes *et al.*, 2019).

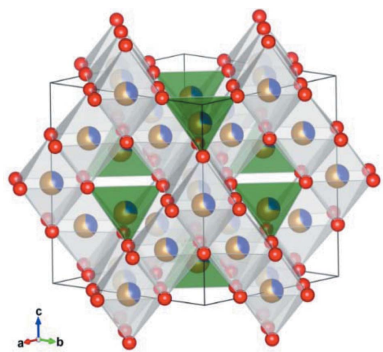

The spinel ferrites, including $CoFe_2O_4$, crystallize in the spinel structure (space group $Fd\bar{3}m$, No. 227), which consists of a cubic close-packed (c.c.p.) structure of $O^{2-}$ anions where 1/8 of the tetrahedral (Td) and 1/2 of the octahedral (Oh) sites are occupied by the cations (Valenzuela, 2012; Suzuki, 2001; Ferreira *et al.*, 2003). The basic formula of spinel structured ferrites is $AB_2O_4$, where $A$ and $B$ represent a divalent metal (*e.g.* $Fe^{2+}$, $Co^{2+}$, $Zn^{2+}$) and trivalent metal ion (*i.e.* $Fe^{3+}$), respectively (Hou *et al.*, 2010; Morrish, 2001) (see Fig. 1). In the following, the metallic cations $A$ and $B$ will be described by $A^{2+}$ and $B^{3+}$, respectively. The unit cell is composed of 32 $O^{2-}$ anions, with 8 $A^{2+}$ and 16 $B^{3+}$ cations occupying the interstitial sites (Mathew & Juang, 2007; Valenzuela, 2012; Na *et al.*, 1993; Morrish, 2001; Gorter, 1955). Classically, the distribution of the $A^{2+}$ and $B^{3+}$ cations on the Td and Oh sites gives rise to either a normal spinel, $(A^{2+})^{Td}[B^{3+}]^{Oh}_2O_4$, or an inverse spinel, $(B^{3+})^{Td}[A^{2+}B^{3+}]^{Oh}O_4$ (Mathew & Juang, 2007; Valenzuela, 2012; Morrish, 2001; Gorter, 1955; Hill *et al.*, 1979). However, the structure can also be partially inverse, $(A^{2+}_{1-x}B^{3+}_{x})^{Td}$-$[A^{2+}_{x}B^{3+}_{2-x}]^{Oh}O_4$, with a fraction ($x$) of the $A^{2+}$ ions occupying the octahedral site. The fraction $x$ is often called the inversion degree (Mathew & Juang, 2007; Valenzuela, 2012; Hou *et al.*, 2010; Gorter, 1955; Hill *et al.*, 1979). A ferrimagnetic structure is generally formed by the opposite alignment of the magnetic moments between the Oh and Td sites, which are coupled through the oxygen atoms by super-exchange interactions (Kim *et al.*, 2001; Sawatzky *et al.*, 1968). As a result, the cation distribution within the sites of the spinel ferrite structure can affect the attained magnetic properties (Coey, 2010*b*). For $CoFe_2O_4$, the different number of unpaired electrons in $Co^{2+}$ (3 e$^-$) and $Fe^{3+}$ (5 e$^-$) makes it a tuneable magnetic system (Mathew & Juang, 2007; Aghavnian *et al.*, 2015; Hou *et al.*, 2010). It should be noted that unpaired electrons may in reality contribute differently to the magnetic moment due to spin and orbit moment contributions. For example, Co has a theoretical 0 K ($-273^{\circ}$C) spin magnetic moment of 3.87 $\mu_B$ (Gorter, 1955; West, 2014); when including the orbit moment the atomic magnetic dipole moment in theory increases to 5.20 $\mu_B$ (West, 2014), while the observed moment is typically found to be between 4.1 and 5.2 $\mu_B$ (West, 2014). The orbit moment contribution explains the large magnetocrystalline anisotropy of cobalt ferrite (Chikazumi *et al.*, 1997). For $Fe^{3+}$, there is no orbital moment; therefore the atomic magnetic dipole moment is the spin-only moment and it is equal to 5.92 $\mu_B$ at 0 K (West, 2014; Gorter, 1955). In reality, the observed atomic magnetic dipole moments in the spinel structures are lower than those expected from theoretical calculations (Chikazumi *et al.*, 1997; West, 2014).

Bulk $CoFe_2O_4$ has an inverse spinel ferrite structure (Na *et al.*, 1993; Hou *et al.*, 2010); however, the inversion degree in nanosized $CoFe_2O_4$ particles has been observed to differ from that of the bulk structure (Andersen *et al.*, 2019; Andersen, Saura-Múzquiz *et al.*, 2018; Rani *et al.*, 2018; Freire *et al.*, 2021). Notably, various inversion degrees have been reported for different synthesis methods and for different heating rates used in the synthesis (Stingaciu *et al.*, 2017; Ferreira *et al.*, 2003; Hou *et al.*, 2010; Sawatzky *et al.*, 1968).

In the literature, the structural characterization of $CoFe_2O_4$ is predominantly carried out by conventional Cu-source laboratory powder X-ray diffraction (PXRD) as this technique is readily available, plus it has higher intensity compared with other anode materials and, for most materials, provides

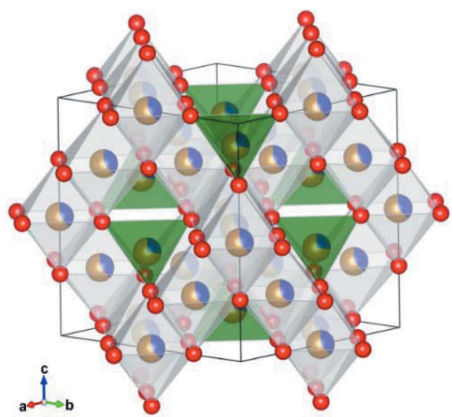

**Figure 1**
Spinel structure of $CoFe_2O_4$. Brown–blue atoms represent the proportion of Fe and Co in the structure, with Fe as brown atoms and Co as blue ones. Red atoms show the position of O. Td sites are shown by the green tetrahedra and Oh sites by white octahedra. This figure was made using the *VESTA* software (Momma & Izumi, 2011).

**Table 1**
Comparison between the resonant scattering terms of $^{8}$O, $^{26}$Fe and $^{27}$Co for Cu $K\alpha_{1,2}$ and Co $K\alpha_{1,2}$ radiation sources, as well as their neutron scattering lengths, $b_j$.

The Fe/Co contrast at $Q = 0$ was calculated for resonant scattering terms and for the neutron scattering length, ignoring the imaginary term. The atomic form factor is given by $f(Q = 0, \lambda)$, $f_0$ corresponds to the Thomson scattering, while $f'$ and $f''$ are the real and imaginary resonant scattering terms, respectively. The resonant scattering values are extracted from Creagh (2004) and the $b_j$ values from Sears (1992).

| | | Co $K\alpha_{1,2}$ ($\lambda$ = 1.79 Å) | Cu $K\alpha_{1,2}$ ($\lambda$ = 1.54 Å) | Neutron scattering length $b_j$ ($10^{-15}$ m) |
|---|---|---|---|---|
| $^{8}$O | $f_{0,O^{2-}}(Q = 0) =$ | 10 | | 5.803 (4) |
| | $f' =$ | +0.0630 | +0.0492 | |
| | $f'' =$ | +0.0440 | +0.0322 | |
| $^{26}$Fe | $f_{0,Fe^{3+}}(Q = 0) =$ | 23 | | 9.45 (2) |
| | $f' =$ | −3.3307 | −1.1336 | |
| | $f'' =$ | +0.4901 | +3.1974 | |
| $^{27}$Co | $f_{0,Co^{2+}}(Q = 0) =$ | 25 | | 2.49 (2) |
| | $f' =$ | −2.0230 | −2.3653 | |
| | $f'' =$ | +0.5731 | +3.6143 | |
| Fe/Co contrast ($Q = 0$) | $\dfrac{f_{Co^{2+}}(0, \lambda) - f_{Fe^{3+}}(0, \lambda)}{f_{Fe^{3+}}(0, \lambda)}$ | 16.8% | 3.5% | $\left\|\dfrac{b_{Co} - b_{Fe}}{b_{Fe}}\right\| = 74\%$ |

The cation distribution within spinel ferrite structures can be investigated by a multitude of local spectroscopy techniques, such as Mössbauer spectroscopy (Smith *et al.*, 1978; Murray & Linnett, 1976*a*,*b*), X-ray absorption spectroscopy with extended X-ray absorption fine structure (Yang *et al.*, 2005; Calvin *et al.*, 2002), X-ray absorption near-edge structure analysis (Nakashima *et al.*, 2007) or Raman spectroscopy (Chandramohan *et al.*, 2011). The site occupation fractions of $Co^{2+}$ and $Fe^{3+}$ can also be extracted by structural modelling of NPD data because of the large scattering length ($b_j$) difference between the two elements (see Table 1) (Sears, 1992). In addition, neutron diffraction data contain information about the magnetic structure of the sample. Notably, a more accurate and robust description of the atomic structure can be obtained by conducting a joint structural modelling of multiple diffraction patterns from different radiation sources.

The present study is based on a compilation of data from three of our earlier studies on $CoFe_2O_4$ nanoparticles (Granados-Miralles *et al.*, 2018; Ahlburg *et al.*, 2020; Stingaciu *et al.*, 2017), as well as previously unpublished data. The X-ray and neutron diffraction patterns stem from several different $CoFe_2O_4$ samples and have been collected using both Co and Cu in-house X-ray diffractometers, as well as various NPD instruments. Here, the data have been re-analysed, yielding a detailed study of the $CoFe_2O_4$ nanoparticle structures and the effect of different modelling parameters on the structural analysis. The study is divided into the following four parts:

Part 1, reliable extraction of Fe/Co occupancies in $CoFe_2O_4$. This work is based on a study by Granados-Miralles *et al.* (2018). Two major points are investigated in this refinement study. (*a*) Pattern weighting influence: investigation of the effect of using an equal weighting (Ew) for each pattern in the joint refinement compared with a weighting based on information available, *i.e.* the number of peaks available for each individual pattern (Iw), and with an arbitrary weighting scheme (Aw). (*b*) Combining different patterns: comparison of the influence of carrying out the joint structural modelling of different combinations of available data from three diffraction sources, *i.e.* Cu $K\alpha_{1,2}$ and Co $K\alpha_{1,2}$ PXRD data and NPD data from the DMC instrument at SINQ (Switzerland). Both the crystal and magnetic structures are analysed and compared, with emphasis on determining which combination gives the most reliable and accurate description.

Part 2, reproducibility of synthesis and refinements. This part investigates the reproducibility of the synthesis using data from samples prepared under identical conditions from a study by Ahlburg *et al.* (2020). Diffraction data have been measured multiple times on each individual sample as well as equivalently prepared samples using identical conditions at reliable structural information. However, in the case of $CoFe_2O_4$ it is challenging to reliably determine the site occupancies of $Co^{2+}$ and $Fe^{3+}$ due to their almost identical scattering power. Another point to consider is that Cu $K\alpha_{1,2}$ radiation ($\lambda_{CuK\alpha}$ = 1.54 Å, $E_{CuK\alpha}$ = 8.05 keV) (Creagh, 2004) has an energy just above the absorption edges of the $K$ level of Co (7.70 keV) and Fe (7.11 keV) (Bearden & Burr, 1967). Due to absorption, the beam will only scatter from a limited volume of the sample and the scattered beam will contain a strong fluorescence signal, which increases the background signal in a diffraction experiment (Drever & Fitzgerald, 1970). In order to circumvent this issue, diffracted beam monochromators (Drever & Fitzgerald, 1970; Strong & Kaplow, 1966; Fransen, 2004), and/or modern detectors equipped with energy discrimination (Bunaciu *et al.*, 2015), can be used to remove the fluorescence background signal. Another way to completely avoid fluorescence is to change the incident beam energy by using either synchrotron radiation with a tuneable wavelength, a different X-ray source (*e.g.* Co source, $\lambda_{CoK\alpha}$ = 1.79 Å, $E_{CoK\alpha}$ = 6.93 keV) (Creagh, 2004) or neutron powder diffraction (NPD). The energy of the Co $K\alpha$ source lies below the absorption edge of Fe and Co, and thus fluorescence can be avoided and a larger volume of cobalt ferrite sample can be probed. In addition, changing the incident beam energy can provide greater contrast between Co and Fe due to changes in the energy-dependent resonant scattering terms (Waseda *et al.*, 1995). The atomic form factor is given by $f(Q, \lambda) = f_0(Q) + f'(\lambda) + if''(\lambda)$, where $f_0$ is the Thomson scattering, which is proportional to the number of electrons ($Z$) around the atom, while $f'$ and $f''$ are the energy-dependent resonant scattering terms (Creagh, 2004). Here, $Q$ represents the scattering vector magnitude and $\lambda$ the wavelength of the incident X-rays. At $Q = 0$ the atomic form factor equals $Z$, plus the resonance terms. Table 1 shows the different values of the scattering terms as a function of the X-ray source used, for O, Fe and Co atoms, as well as their neutron scattering lengths $b_j$.

Table 2
Overview of the presented samples and references to papers where some of the results have previously been presented.

\* indicates this work, x data published in the cited references.

|  | Sample | Temperature (°C) | Time (h) | Heating rate (°C s$^{-1}$) | VSM | PXRD | NPD | Occupancy reported | Refined |
|---|---|---|---|---|---|---|---|---|---|
| DMC | CoFe$_2$O$_4$ (Granados-Miralles et al., 2018) | 240 | 2 | 0.15 | x | x | x | Yes | * |
| DMC | A1-3* | 240 | 2 | 0.15 | x | x | x | No | * |
|  | B1-2* | 240 | 2 | 0.15 | x | x | x | No | * |
|  | C1-3 (Ahlburg et al., 2020) | 240 | 2 | 0.15 | x | x | x | Yes | * |
| PUS | FR220 (Stingaciu et al., 2017) | 220 | 3 × 10$^{-3}$ | 500 | x | x | * | * | * |
|  | FR320 (Stingaciu et al., 2017) | 320 | 3 × 10$^{-3}$ | 500 | x | x | * | * | * |
|  | SR240 (Stingaciu et al., 2017) | 240 | 1/3 | 25 | x | x | * | * | * |
|  | AC240 (Stingaciu et al., 2017) | 240 | 1 | 0.15 | x | x | * | * | * |
| HRPT | Co(Ac)$_2$* | 200 | 1 | 0.15 | * | * | * | * | * |
|  | Co(NO$_3$)$_2$* | 200 | 1 | 0.15 | * | * | * | * | * |
|  | CoCl$_2$* | 200 | 1 | 0.15 | * | * | * | * | * |

the DMC instrument. This allows us to investigate the reproducibility of both the synthesis method and the data acquisition and refinement procedure.

Part 3, effect of different synthesis approaches. Here, the effect of different hydrothermal synthesis approaches on the structural and magnetic properties of CoFe$_2$O$_4$ is investigated. This part is divided into two subsections. (a) Different hydrothermal reactors: an extensive study prepared CoFe$_2$O$_4$ using different hydrothermal reactors (autoclave, spiral reactor and continuous flow reactor) conducted by Stingaciu et al. (2017). From this study four samples were investigated using NPD. (b) Different Co salts: comparison of the structural consequences of employing different Co salt precursors [Co acetate, CoCl$_2$ and Co(NO$_3$)$_2$] in the synthesis of CoFe$_2$O$_4$.

Part 4, effect of synthesis conditions. This part compares and discusses four samples presented in the paper that have been synthesized using the autoclave reactor. The chemicals and synthesis parameters used are considered in the comparison of their crystal and magnetic structures to deduce possible trends.

## 2. Experimental

### 2.1. Sample preparation

The CoFe$_2$O$_4$ samples used for these investigations were all prepared using variations of the same hydrothermal method. An overview of all samples is given in Table 2 and the sample preparation is described in the following paragraphs.

For part 1, the CoFe$_2$O$_4$ powder was prepared following the method described by Granados-Miralles et al. (2018). Aqueous solutions of 2.0 M Co(NO$_3$)$_2$·6H$_2$O and 2.0 M Fe(NO$_3$)$_3$·9H$_2$O were mixed in the ratio 1:2, and precipitates were formed by adding 16 M NaOH in a ratio of 2:1 with respect to OH$^-$:NO$_3^-$. All reagents used were technical grade with purity >98% from Sigma–Aldrich. The obtained precursor was diluted with deionized water to a metal concentration of 0.45 M. The formed precipitate was stirred for 30 min and 80 ml were transferred to a 170 ml Teflon-lined steel autoclave. The autoclave was placed in a preheated convection oven at 240°C, held for 2 h and subsequently cooled in air. Finally, the sample was washed with deionized water several times to remove the counter-ions and neutralize the supernatant, and then dried in a vacuum oven for 4 h at 70°C. The same cleaning method was also applied in the following unless stated otherwise.

For part 2, the CoFe$_2$O$_4$ powders were prepared using the method described by Ahlburg et al. (2020). Two autoclave syntheses similar to the method described in part 1 were used with the exception of the usage of higher reagent concentrations of 3.0 M Co(NO$_3$)$_2$·6H$_2$O and 2.3 M Fe(NO$_3$)$_3$·9H$_2$O, but still with a 1:2 Co:Fe ratio and an OH$^-$:NO$_3^-$ molar ratio of 1.25:1 when adding NaOH. The same heating process was used as for part (1), but the samples were dried in a vacuum oven for 24 h at 50°C.

For part 3, three different hydrothermal reactors (autoclave, spiral reactor and flow reactor) with different heating rates were used. In addition, three different Co salts were used to prepare the precursors for the autoclave synthesis. The details are as follows:

(a) Different hydrothermal reactors. In this part, the precursor consisted of 2.0 M Co(NO$_3$)$_2$·6H$_2$O, 2.0 M Fe(NO$_3$)$_3$·9H$_2$O and 16.0 M NaOH with the molar ratios of Co:Fe and OH$^-$:NO$_3^-$ being 1:2 and 2:1, respectively. The synthesis was carried out using the following reactors and specific conditions:

Autoclave (AC): the synthesis was carried out in a conventional Teflon-lined steel AC (170 ml filled with 80 ml precursor) over a period of 1 h at 240°C. The corresponding sample name is AC240.

Spiral reactor (SR): the synthesis was carried out in a custom-built Swagelok steel tube spiral reactor (Granados-Miralles et al., 2016). The precursor solution was diluted to a metal concentration of 0.45 M, before being loaded into the 1/16 inch (1.59 mm) steel tube spiral. The reactor was pressurized to 210 bar (1 bar = 100 kPa) using a high-performance liquid chromatography pump. The precursor was subsequently submerged in a hot oil bath at 240°C for 20 min (Granados-Miralles et al., 2016). The corresponding sample name is SR240.

**Table 3**
Wavelength ($\lambda$), angular coverage ($2\theta$), step resolution of the detector ($\Delta 2\theta$), $Q$ coverage and $Q$ step size ($\Delta Q$) of the two X-ray instruments and the three neutron powder diffractometers.

The number in parentheses refers to the relevant part of the study. Both DMC and HRPT used a radial collimator to reduce the background signal from the surroundings.

| Source | $\lambda$ (Å) | Detector | Monochromator | $2\theta$ range (°) | $\Delta 2\theta$ (°) | $Q$ coverage (Å$^{-1}$) | $\Delta Q$ (Å$^{-1}$) |
|---|---|---|---|---|---|---|---|
| Cu $K\alpha_{1,2}$ | 1.54 | Dtex/Ultra | Diffracted beam monochromator | (1) 16–108 | 0.015 | (1) 1.13–6.60 | 7.12 × 10$^{-4}$ |
| | | | | (3a) 14–135 | 0.015 | (3a) 1.00–7.54 | 7.12 × 10$^{-4}$ |
| | | | | (3b) 14–120 | 0.015 | (3b) 1.00–7.06 | 7.12 × 10$^{-4}$ |
| Co $K\alpha_{1,2}$ | 1.79 | Dtex/Ultra | None | (1) 17–140 | 0.015 | (1) 1.04–6.60 | 6.13 × 10$^{-4}$ |
| | | | | (2) 15–120 | 0.015 | (2) 0.92–6.08 | 6.13 × 10$^{-4}$ |
| DMC | 2.46 | Multiwire detector | Pyrolytic graphite (002), vertically focusing | (1) 11–92.7 | 0.1 | (1) 0.49–3.70 | 4.46 × 10$^{-3}$ |
| | | | | (2) 12.8–92.9 | 0.1 | (2) 0.57–3.70 | 4.46 × 10$^{-3}$ |
| HRPT | 1.49 | Multiwire detector | Ge monochromator, vertically focusing | (3b) 3.8–164.75 | 0.05 | (3b) 0.28–8.36 | 3.68 × 10$^{-3}$ |
| PUS | 1.55 | 14 individual $^3$He counters | Ge monochromator (511) | (3a) 10.3–129.95 | 0.05 | (3a) 0.73–7.35 | 3.54 × 10$^{-3}$ |

Flow reactor (FR): the synthesis was carried out using a single-stage continuous solvothermal flow reactor (Hellstern *et al.*, 2015; Søndergaard *et al.*, 2011). The precursor solution was diluted to a concentration of 0.05 *M* to avoid clogging the tubing of the FR before being transferred to a 200 ml injector. The system was pressurized to 250 bar and the precursor and preheated solvent (water) were pumped continuously at flow rates of 5 and 15 ml min$^{-1}$, respectively (Stingaciu *et al.*, 2017). Two samples were prepared, at 220°C (sample name: FR220) and at 320°C (sample name: FR320).

(*b*) Different cobalt salts. The syntheses using different cobalt sources required slight modifications compared with the previous synthesis, due to the reduced aqueous solubilities of the used cobalt salts. The salts used were cobalt(II) chloride (CoCl$_2$), cobalt(II) nitrate [Co(NO$_3$)$_2$] and cobalt(II) acetate [Co(CH$_3$COOH)$_2$]. Here, 20 ml of Fe(NO$_3$)$_3$·9H$_2$O (2.0 *M*) were mixed with 20 ml of either CoCl$_2$, Co(NO$_3$)$_2$ or Co(CH$_3$COOH)$_2$ (1.0 *M*) to obtain a Co:Fe molar ratio of 1:2. After thorough mixing, 30 ml of NaOH (12 *M*) was added, giving an OH$^-$:Cl$^-$/NO$_3^-$/CH$_3$COOH$^-$ ratio of 2.25:1. The precursor was sealed in a 170 ml Teflon-lined steel autoclave and placed in a convection oven at 200°C for 1 h. In the following, cobalt(II) acetate will be abbreviated as Co(Ac)$_2$.

### 2.2. Structural characterization

Several different X-ray and neutron powder diffraction instruments were used for the structural characterization. An overview of the employed instruments is given in Table 3 and they are described below.

Room-temperature PXRD patterns were collected on two in-house Rigaku SmartLab powder X-ray diffractometers (Rigaku, Japan), one equipped with a Cu source (Cu $K\alpha_{1,2}$; $\lambda_1$ = 1.54059 Å; $\lambda_2$ = 1.54441 Å) and the other equipped with a Co source (Co $K\alpha_{1,2}$; $\lambda_1$ = 1.78892 Å; $\lambda_2$ = 1.79278 Å), both in Bragg–Brentano geometry. The powder diffraction patterns were collected using a Dtex/Ultra detector in fluorescence suppression mode and a diffracted beam monochromator was placed in front of the detector on the Cu-source instrument. The data collection is summarized in Table 3.

The NPD patterns were measured at room temperature at two different neutron sources: at the Swiss spallation neutron source (SINQ) (Blau *et al.*, 2009; Allenspach, 2000), Paul Scherrer Institute (PSI), in Switzerland, and at the Institute for Energy Technology (IFE), Kjeller, Norway, using different instrumentation. At SINQ two diffractometers were used: DMC, the cold neutron powder diffractometer (https://www.psi.ch/en/sinq/dmc), and HRPT, the high-resolution powder diffractometer for thermal neutrons (https://www.psi.ch/en/sinq/hrpt) (Fischer *et al.*, 2000). At IFE the PUS diffractometer was used (Hauback *et al.*, 2000).

The instrumental contributions to the peak broadening of the collected patterns were determined from data collected on standard reference materials at equivalent instrumental configurations. An LaB$_6$ NIST 660B standard was used for the in-house PXRD experiments (Black *et al.*, 2011), while an Na$_2$Ca$_3$Al$_2$F$_{14}$ standard was used for the NPD data at DMC and HRPT (Courbion & Ferey, 1988), and a CeO$_2$ standard was used at PUS.

### 2.3. Vibrating sample magnetometry

The magnetic properties of the samples were characterized using a Physical Property Measurement System (PPMS), from Quantum Design, equipped with a vibrating sample magnetometer (VSM). The measurements were performed on cold-pressed cylindrical pellets with a diameter of 3 mm, gently compacted using a hand-held press. The field-dependent magnetization curves (expressed in A m$^2$ kg$^{-1}$) of the pellets were measured by scanning the externally applied field, $H_{app}$, between ±3 T at 27°C. The saturation magnetization $M_{sat}^{VSM}$ was obtained by applying the law of approach to saturation to the data (Zhang *et al.*, 2010).

### 2.4. Structural refinement

The structural analyses were carried out using the *FullProf Suite* software (Rodríguez-Carvajal, 1993). The crystalline CoFe$_2$O$_4$ phase was in all cases described in the space group $Fd\bar{3}m$ with the Laue class $m\bar{3}m$. As described by Andersen *et al.* (2019), the magnetic structure was implemented as an additional magnetic phase with the lowest-symmetry space group of the corresponding centring, i.e. $F\bar{1}$. The special positions of $Fe^{3+}$ and $Co^{2+}$ cations were specified, and the first 24 symmetry operations of the $Fd\bar{3}m$ space group were provided to generate all atomic magnetic moments. Additional scattering factors (resonant terms) were added for each of the X-ray sources (see Table 1) to distinguish Fe and Co.

The peak profile parameters were described using the Thompson–Cox–Hastings (TCH) pseudo-Voigt function (Thompson et al., 1987), while the crystallites were assumed to be strain free. The crystallite size was extracted from the Lorentzian isotropic size parameter, $Y$, which was constrained between data sets through appropriate consideration of the wavelength. This constraining process is explained in detail in the supporting information.

The scale factors for X-ray data sets were refined individually, in contrast to the scale factors for nuclear and magnetic phase for the neutron data, which were kept identical as they belong to the same data set. The backgrounds were described by Chebyshev polynomials, with a maximum of six refinable coefficients (Bck_0, 1, 2, 3, 4 and 5). The wavelength for the neutron data was refined, allowing a single common unit-cell parameter to be refined for all the diffraction patterns, along with the zero point. The fractional coordinates of oxygen $(x, x, x)$ were also refined. The site occupation fraction (Occ) was refined with a constraint, as were the atomic displacement parameters (ADPs) and the atomic magnetic dipolar moment $(R_x)$. The refinement of these three parameters is described below.

For low-$Q$-coverage neutron data ($Q_{max} < 4$ Å, i.e. DMC), the ADPs were refined as one common isotropic displacement parameter ($B_{ov}$) for all atoms. We set the Td and Oh sites to be fully occupied while refining the relative Co and Fe occupancies with the total Co and Fe content constrained to the nominal Co:Fe ratio of 1:2. This means that $CoFe_2O_4$ was refined as $(Co^{2+}_{1-x} Fe^{3+}_x)^{Td}[Co^{2+}_x Fe^{3+}_{2-x}]^{Oh}O_4$, with $x$ being the inversion degree, where $x = 0$ is normal spinel and $x = 1$ is fully inverse.

The magnetic structure was constrained on the basis of the number of unpaired electrons in $Fe^{3+}$ ($3d^5$) and $Co^{2+}$ ($3d^7$). Thus, the magnetic moments of $Co^{2+}$ and $Fe^{3+}$ were refined in a 3:5 ratio, without taking the Co orbital contribution into account, i.e. assuming the orbit moment is quenched. The moment was refined along the (100) direction using the $R_x$ parameter, where $R_x$(Oh) was chosen to be positive and $R_x$(Td) negative. The effect of including the Co orbital contribution to the magnetic moment was investigated and is reported in the supporting information. It is not straightforward to determine which model is better suited to describe the magnetic structure of $CoFe_2O_4$. According to the VSM measurement, a better match between the experimental data and the refined NPD values was found when the orbit moment for Co was quenched.

For the uncertainties, the standard deviation provided by the *FullProf Suite* software was used except for the crystallite size. Here, the uncertainty was chosen to be equal to the unit-cell length, i.e. 0.8 nm, instead of the very low (0.001–0.01 nm) mathematical uncertainties provided by *FullProf*. Given that the unit cell constitutes the smallest crystal unit, we consider using the cell parameter as the uncertainty for the average crystallite size to be more appropriate from a physical perspective. The uncertainty of the calculated net intrinsic magnetization ($M^{neutron}$) is based on the standard deviation of each element constituting the calculation.

## 3. Results and discussion

### 3.1. Part 1, reliable extraction of Fe/Co occupancies in $CoFe_2O_4$

(*a*) Pattern weighting influence. When jointly refining multiple diffraction patterns, the weighting between data sets must be carefully considered as this may significantly influence the outcome of the refinement. As noted by both Deutsch et al. (2012) and Coppens et al. (1981), the weighting scheme constitutes one of the main issues in joint refinement strategies. Yet, this remains a topic of much debate, with the literature providing no definitive answer, and thus the weighting schemes are generally somewhat subjectively chosen. Two weighting schemes seem to be prevalent: (i) a weighting that minimizes the sum of the goodness of fit ($R$ factors and $\chi^2$) of each pattern (Deutsch et al., 2012; Coppens et al., 1981; Duckworth et al., 1969; Kibalin et al., 2017) and (ii) a weighting scheme that minimizes the sum of the goodness of fit normalized per data number for each pattern, by using log functions (Deutsch et al., 2012; Kibalin et al., 2017; Gillet & Becker, 2004). The scheme adopted by the *FullProf Suite* software appears to correspond to (i), by considering the residual function defined as $\chi^2 = \sum_{i=1}^{n} w_i[y_i - y_{c,i}(\alpha)]^2$, where $[y_i - y_{c,i}(\alpha)]^2$ represents the difference between the experimental and the calculated patterns, while the statistical weight $w_i$ is defined as the inverse of the squared variance of the observed pattern, $\sigma_i$. In our refinement, the minimized function of the whole refinement can be defined as $\chi^2_{tot} = \sum_{P=1}^{n} w_P \chi^2_P$ with $w_P$ and $\chi^2_P$ the weight and chi-squared of the individual pattern $P$, respectively. Therefore step size, uncertainties and intensities are to be considered if a proper weighting of different data is to be carried out.

Here, we have considered three weighting schemes and made a simple comparative study of the influence of the pattern weighting on the refinement of atomic and microstructural parameters of cobalt ferrite. In conclusion, we found that the three tested weighting models gave very similar results and/or are within the uncertainty of each other, indicating that the weighting scheme does not hugely affect the refinements of the present data. Therefore, in the following, we have used the standard *FullProf* weighting model using equivalent weight of all patterns. The study of the three weighting schemes can be found in the supporting information.

(*b*) Combining different patterns. Joint refinements of different combinations of the PXRD (Co and/or Cu) and NPD data are evaluated, with special emphasis on investigating the reliability/consistency of the Co/Fe occupation fractions

**Table 4**
List of refined parameters for the seven different combinations of data sets.

In all combinations the wavelength of the neutron data was refined, except in (v) (DMC) where it was fixed to 2.45948 Å, based on the refined value of (i) (DMC/Co/Cu). The saturation magnetization extracted from a VSM measurement ($M_{sat}^{VSM}$) is tabulated along with the formula unit magnetic moment ($m_{f.u.}$), as well as the net intrinsic crystallographic magnetization ($M^{neutron}$) of $CoFe_2O_4$, calculated from the refined magnetic structure. The numbers in parentheses represent the uncertainties of the *FullProf Suite* software, except for $x$ and $M^{neutron}$ where the uncertainties were calculated by the propagation of errors. The number of reflections is written as #reflections in the table. '–' means there is no value attributed to the refined parameter. The diffraction data were previously published by Granados-Miralles *et al.* (2018), but all refinements were redone for this work.

| | (i) | (ii) | (iii) | (iv) | (v) | (vi) | (vii) |
|---|---|---|---|---|---|---|---|
| | DMC/Co/Cu | DMC/Co | DMC/Cu | Co/Cu | DMC | Co | Cu |
| Unit cell (Å) | 8.3892 (1) | 8.3890 (1) | 8.3891 (3) | 8.3890 (1) | 8.3889 (1) | 8.3890 (1) | 8.3891 (1) |
| Crystallite size (nm) | 13.2 (8) | 13.3 (8) | 13.1 (8) | 13.3 (8) | 12.4 (8) | 13.3 (8) | 13.8 (8) |
| $x$ (O) | 0.2425 (1) | 0.2428 (1) | 0.2411 (3) | 0.2430 (1) | 0.2408 (4) | 0.2429 (1) | 0.2425 (2) |
| $B_{ov}$ (Å$^2$) | 1.07 (1) | 1.22 (1) | 0.89 (4) | 1.22 (1) | 0.13 (8) | 1.20 (1) | 1.05 (2) |
| Occ($Co^{2+}$)$^{Td}$ (%) | 24 (1) | 24 (1) | 29 (1) | 20 (1) | 28 (1) | 21 (1) | −47 (8) |
| Occ($Fe^{3+}$)$^{Td}$ (%) | 76 (1) | 76 (1) | 71 (1) | 80 (2) | 72 (1) | 79 (1) | 147 (25) |
| Occ($Co^{2+}$)$^{Oh}$ (%) | 38 (1) | 38 (1) | 35 (1) | 40 (1) | 36 (1) | 40 (1) | 74 (13) |
| Occ($Fe^{3+}$)$^{Oh}$ (%) | 62 (1) | 62 (1) | 65 (1) | 60 (1) | 64 (1) | 60 (1) | 26 (4) |
| $x$ | 0.76 (2) | 0.76 (2) | 0.71 (1) | 0.80 (2) | 0.72 (1) | 0.79 (2) | 1.47 (35) |
| $R_x(Co^{2+})^{Oh}$ ($\mu_B$) | 2.33 (1) | 2.33 (1) | 2.31 (1) | – | 2.38 (2) | – | – |
| $R_x(Fe^{3+})^{Oh}$ ($\mu_B$) | 3.89 (2) | 3.88 (2) | 3.86 (2) | – | 3.96 (3) | – | – |
| $m_{f.u.}$ ($\mu_B$ f.u.$^{-1}$) | 3.1 (1) | 3.1 (1) | 3.2 (1) | – | 3.3 (1) | – | – |
| $M^{neutron}$ (A m$^2$ kg$^{-1}$) | 73 (2) | 73 (3) | 77 (2) | – | 78 (2) | – | – |
| $M_{sat}^{VSM}$ (A m$^2$ kg$^{-1}$) | 73.5 (2) | 73.5 (2) | 73.5 (2) | 73.5 (2) | 73.5 (2) | 73.5 (2) | 73.5 (2) |
| $R_{wp}$ (%) | 6.0/7.0/15.3 | 6.2/6.6 | 5.0/16.0 | 6.6/16.0 | 4.4 | 6.7 | 15.7 |
| $\chi^2$ | 4.7/4.8/1.0 | 5.0/5.0 | 3.3/1.0 | 5.0/1.0 | 2.5 | 5.1 | 0.9 |
| $R_{Bragg}$ (%) | 2.5/2.8/5.1 | 2.5/2.5 | 1.1/7.2 | 2.4/8.7 | 0.5 | 2.4 | 6.9 |
| $R_{mag}$ (%) | 0.95/–/– | 1.0/– | 0.8/– | –/– | 0.36 | – | – |
| #reflections | 6/18/18 | 6/18 | 6/18 | 18/18 | 6 | 18 | 18 |

obtained from the refinements. Refinements of seven different combinations of the three patterns were tested using equal weighting: (i) DMC/Co/Cu, (ii) DMC/Co, (iii) DMC/Cu, (iv) Co/Cu, (v) DMC, (vi) Co and (vii) Cu. Table 4 gives an overview of the extracted refinement parameters. Each individual pattern is described in the supporting information. The refined powder diffraction patterns from combination (i) (DMC/Co/Cu) can be seen in Fig. 2, while the six remaining combinations of patterns are reported in Fig. S2 (see the supporting information).

The in-depth investigation of the different combinations reveals the models that best describe the crystal and magnetic structure, as well as the Co/Fe occupation fraction. The different models are discussed in the following.

Using only the data set (vii) (Cu) is not a good choice for refining the Co/Fe site occupation fractions as unphysical values are obtained. This is not surprising because the Cu data have almost no scattering contrast between Co and Fe. The neutron data with limited $Q$ range alone [model (v)] are also not a good choice, as the refinement yields an unrealistic description of the ADPs due to the limited number of Bragg peaks (Andersen, Bøjesen *et al.*, 2018). It is perhaps more surprising that the data set (vi) (Co) actually performs very well compared with the combined X-ray and neutron models. If information on Fe/Co occupation is sought, using data set (vi) (Co) alone saves time with regard to data collection. The refinement of data set (vi) (Co) is close to identical to combinations (i) (DMC/Cu/Co) and (ii) (DMC/Co). This contradicts the common notion that in-house PXRD data alone cannot be used to extract reasonable occupation fractions for neighbouring elements. When using Co radiation, it is the resonant scattering terms ($f'$, $f''$) of Fe and Co that enhance the contrast between the elements.

Also the joint refinement of Cu and DMC [data set (iii)] provides a good structural description, making this a feasible data set combination for future diffraction experiments.

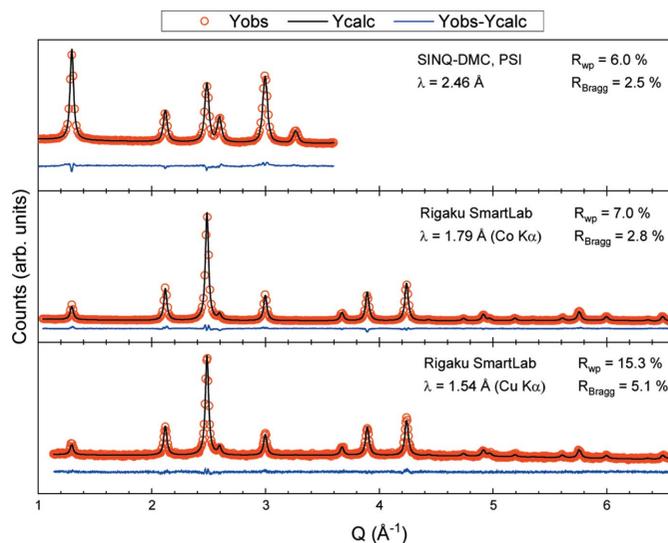

**Figure 2**
Combined Rietveld refinement of $CoFe_2O_4$ diffraction patterns obtained using neutrons (DMC), Co and Cu as radiation sources. The data are shown as red circles, the refined model as a black line and the residual as a blue line. Weighted profile, $R_{wp}$, and Bragg, $R_{Bragg}$, $R$ factors are indicated for each diffraction pattern. For visualization purposes, a specified frequency of data points has been selected: frequencies of 3, 20 and 10 points have been drawn for DMC, Co $K\alpha$ and Cu $K\alpha$ patterns, respectively.

**Table 5**
List of refined parameters for samples A, B and C.

The magnetic moment dipole was refined anti-parallel on the Td and Oh sites. Net intrinsic magnetization ($M^{neutron}$) was calculated on the basis of the $R_x$ values obtained from the refinement. The multiple measurements are referred as A1, A2, A3, B1, B2, and C1, C2, C3. The 'Average' column is based on the multiple refined parameters obtained for samples A, B and C. The uncertainties of the average column are based on the standard deviation. The neutron diffraction data are part of an *in situ* study published by Ahlburg *et al.* (2020), but all refinements were redone for this work. All measurements made using model (ii) (DMC/Co).

| | A1 | A2 | A3 | B1 | B2 | C1 | C2 | C3 | Average |
|---|---|---|---|---|---|---|---|---|---|
| Unit cell (Å) | 8.3912 (1) | 8.3912 (1) | 8.3912 (1) | 8.3927 (1) | 8.3927 (1) | 8.3919 (1) | 8.3919 (1) | 8.3919 (1) | 8.3919 (6) |
| Crystallite size (nm) | 13.2 (8) | 13.1 (8) | 13.2 (8) | 12.6 (8) | 12.6 (8) | 13.1 (8) | 13.1 (8) | 13.2 (8) | 13.0 (3) |
| $x$ (O) | 0.2435 (1) | 0.2435 (1) | 0.2435 (1) | 0.2431 (1) | 0.2432 (1) | 0.2434 (1) | 0.2434 (1) | 0.2433 (1) | 0.2434 (2) |
| $B_{ov}$ (Å$^2$) | 1.57 (1) | 1.57 (1) | 1.57 (1) | 1.47 (1) | 1.47 (1) | 1.44 (1) | 1.44 (1) | 1.44 (1) | 1.50 (6) |
| Occ(Co$^{2+}$)$^{Td}$ (%) | 18 (1) | 19 (1) | 18 (1) | 21 (1) | 20 (1) | 19 (1) | 19 (1) | 19 (1) | 19 (1) |
| Occ(Fe$^{3+}$)$^{Td}$ (%) | 82 (2) | 81 (2) | 82 (2) | 79 (2) | 80 (2) | 81 (2) | 81 (2) | 81 (2) | 81 (1) |
| Occ(Co$^{2+}$)$^{Oh}$ (%) | 41 (1) | 41 (1) | 41 (1) | 40 (1) | 40 (1) | 40 (1) | 40 (1) | 40 (1) | 40 (1) |
| Occ(Fe$^{3+}$)$^{Oh}$ (%) | 59 (1) | 59 (1) | 59 (1) | 60 (1) | 60 (1) | 60 (1) | 60 (1) | 60 (1) | 60 (1) |
| $x$ | 0.82 (3) | 0.81 (3) | 0.82 (3) | 0.79 (3) | 0.80 (3) | 0.81 (2) | 0.81 (2) | 0.81 (3) | 0.81 (1) |
| $R_x$(Co$^{2+}$)$^{Oh}$ ($\mu_B$) | 2.37 (3) | 2.42 (2) | 2.43 (3) | 2.42 (2) | 2.59 (4) | 2.70 (6) | 2.70 (6) | 2.56 (5) | 2.5 (1) |
| $R_x$(Fe$^{3+}$)$^{Oh}$ ($\mu_B$) | 3.96 (5) | 4.03 (4) | 4.05 (5) | 4.03 (4) | 4.32 (6) | 4.50 (11) | 4.51 (10) | 4.27 (8) | 4.2 (2) |
| $m_{f.u.}$ ($\mu_B$ f.u.$^{-1}$) | 2.9 (2) | 3.0 (1) | 3.0 (2) | 3.1 (2) | 3.3 (2) | 3.4 (2) | 3.4 (2) | 3.2 (2) | 3.2 (2) |
| $M^{neutron}$ (A m$^2$ kg$^{-1}$) | 70 (4) | 72 (3) | 72 (4) | 73 (4) | 78 (5) | 81 (5) | 81 (5) | 77 (5) | 75 (4) |
| $M^{VSM}_{sat}$ (A m$^2$ kg$^{-1}$) | 72.9 (1) | 72.9 (1) | 72.9 (1) | 73.8 (4) | 73.8 (4) | 73.9 (1) | 73.9 (1) | 73.9 (1) | 73.5 (6) |
| $R_{wp}$ (%) | 14.8/5.3 | 12.3/5.3 | 15.1/5.3 | 12/6.6 | 16.9/6.6 | 20.2/6.1 | 20.1/6.1 | 21.4/6.1 | |
| $\chi^2$ | 1.8/3 | 2.4/3 | 1.7/3 | 1.9/2.3 | 1.4/2.3 | 1.1/4.2 | 1.2/4.2 | 2.5/4.2 | |
| $R_{Bragg}$ (%) | 7.2/2.3 | 6.7/2.4 | 7.1/2.3 | 5.7/2.4 | 4.5/2.4 | 1.6/2 | 3.8/2 | 5.5/2 | |
| $R_{mag}$ (%) | 4.7/– | 4.5/– | 3.8/– | 3.1/– | 4.8/– | 3.3/– | 5.0/– | 6.3/– | |
| #reflections | 6/16 | 6/16 | 6/16 | 6/16 | 6/16 | 6/16 | 6/16 | 6/16 | |

The refinements of data set combinations (iv) (Co/Cu) and (vi) (Co) yield very similar structural descriptions, indicating that, in this case, the inclusion of the Cu pattern does not significantly degrade the accuracy of the results.

In conclusion, four of the examined data combinations can be used to obtain a reliable refinement of Fe/Co occupancies in CoFe$_2$O$_4$: (i) (DMC/Co/Cu), (ii) (DMC/Co), (iii) (DMC/Cu) and (vi) (Co). A neutron diffraction pattern with a larger $Q$ range would also be able to produce reliable refinements of Fe/Co occupancies and the magnetic moment. Despite the limited $Q$ range of the DMC instrument, it was shown that only six peaks in NPD data are sufficient to extract reliable magnetic moments. Combining NPD with PXRD data gives a more robust refinement of the occupation factors as it removes correlations between ADPs and magnetic moments inherent to a single data set. Moreover, the refined formula unit magnetic moment $m_{f.u.}$ value is close to that of the bulk CoFe$_2$O$_4$, and good agreement was seen between the calculated net intrinsic magnetization and the saturation magnetization obtained by VSM measurement. Even if combinations (ii) (DMC/Co) and (iii) (DMC/Cu) are comparable, we recommend using the model involving the Co source (ii) as it also carries information about the Fe/Co occupancies. Combination (i) (DMC/Co/Cu) constituted our benchmark as it has the largest quantity of data; however, the joint refinement of combination (ii) (DMC/Co) gives close to identical values to combination (i) (DMC/Co/Cu), and therefore combination (ii) is favoured as it involves collecting and treating fewer data to obtain the same result. The PXRD data in this case provide better peak resolution and $Q$ coverage, which is essential for describing the microstructure (lattice parameter, crystallite size and thermal vibrations). While it is preferable to have multiple patterns, the study reveals that it is possible to achieve a reasonable refinement of the Fe/Co occupancies using data from a Co X-ray source alone [model (vi)].

### 3.2. Part 2, reproducibility study

This second part of the study investigates the reproducibility of the autoclave synthesis method on the basis of data from three CoFe$_2$O$_4$ nanoparticle samples (A, B and C), reported by Ahlburg *et al.* (2020), synthesized under identical conditions. The reproducibility of the PXRD and NPD measurements, as well as the Rietveld refinements, was investigated. The NPD data were measured three times for the samples A and C, while sample B was measured twice at the DMC instrument. All these multiple measurements were acquired under identical conditions. Table 5 shows the refined structural values obtained for samples A, B, and C, while their diffraction patterns can be found in the supporting information (Figs. S3–S5).

The unit cell, apparent crystallite size (ACS) and oxygen position are all identical within the uncertainties of each repetition and between all samples. Only minute variations were noticed in the unit cell and thermal vibration parameters between the different samples; this can be attributed to a reduced probed sample volume for the $Q$ range exceeding 5 Å$^{-1}$ due to penetration of the sample as revealed by an Al(222) reflection originating from the sample holder (see the supporting information). This could also explain the larger $B_{ov}$ found here compared with values reported in the literature (~0.65 to ~1.05 Å$^2$) (Ferreira *et al.*, 2003; Waseda *et al.*, 1995; Tanaka *et al.*, 2016), and the one found for model (ii) [1.22 (1) Å$^2$, DMC/Co] in part 1(*b*) (see Table 4).

The effect of different thermal vibration values was tested on sample C, by fixing $B_{ov}$ to 1.57 Å$^2$ (the value from sample A). The refinements are shown in the supporting information (Table S5) and named C_BovFIX. Only $R_x$ is affected by changing $B_{ov}$ and the obtained values are well within the uncertainties. The changes in $B_{ov}$ do not even affect the $R$ factors.

The parameters that differ between samples and repetitions are the site occupancy and the magnetic parameters. They are discussed individually below.

Site occupancy. The obtained inversion degree, $x$, is comparable for all data sets and the values range between 0.79 (3) and 0.82 (3). In other words, $x$ is identical between the multiple measurements, but is higher than other reported inversion degrees for CoFe$_2$O$_4$ nanoparticles ($x$ = 0.69–0.79) (Sawatzky et al., 1968; Ferreira et al., 2003; Chandramohan et al., 2011; Tanaka et al., 2016). The values reported here are comparable to those reported by Granados-Miralles et al. (2018) and by Ahlburg et al. (2020), which are based on the same data, but with slight differences in the applied refinement model.

The cation distribution within a spinel ferrite compound is reported to be synthesis dependent (Andersen et al., 2019; Sawatzky et al., 1968; Ferreira et al., 2003; Chandramohan et al., 2011; Moumen et al., 1996). Here it is demonstrated that the autoclave synthesis is robust and the repeated data collection gives identical cation distributions.

Atomic ($R_x$) and formula unit magnetic moment ($m_{f.u.}$). The extracted atomic magnetic dipole moments ($R_x$) reveal small differences between the different multiple measurements of the same sample, and across the batches. Sample A has identical $R_x$ within uncertainties while the variations of the atomic magnetic moment are larger in the case of samples B and C. For sample C3 an additional Bragg peak in the NPD pattern is observed at ~3.1 Å$^{-1}$ – this peak is attributed to Ni(111) from a thermocouple used for monitoring the sample temperature. The Ni peak affects the residual between the experimental and calculated patterns (see Fig. S5), possibly affecting the refinement of the background, which in turn could reduce $R_x$ compared with C1 and C2.

Considering the magnetic moment per formula unit ($m_{f.u.}$), it is found that all samples have similar values [2.9 (2)–3.4 (2) $\mu_B$] within 2$\sigma$ uncertainties, and that the multiple measurements yield the same refined value.

Net intrinsic magnetization ($M^{neutron}$) and saturation magnetization ($M^{VSM}_{sat}$). For sample A, the calculated $M^{neutron}$ values are close to identical between the three measurements, and all values are within the same uncertainty range. Comparison with macroscopic experimental values ($M^{VSM}_{sat}$) reveals good agreement between macroscopic and microscopic magnetization.

Sample C exhibits the highest net magnetization [77 (5)–81 (5) A m$^2$ kg$^{-1}$] as extracted from the occupancies and magnetic moment. The obtained values are higher than the saturation magnetization [73.9 (1) A m$^2$ kg$^{-1}$] measured by VSM. Nonetheless, the relatively large uncertainties on the extracted magnetization show that the obtained $M^{neutron}$ is within 2$\sigma$ of $M^{VSM}$.

Comparing $M^{VSM}_{sat}$ for the three samples shows that they have close to identical saturation magnetization [72.9 (1)–73.9 (1) A m$^2$ kg$^{-1}$], meaning that the AC synthesis yields reproducible results. Sample C exhibits an $M^{neutron}$ higher than the saturation magnetization from $M^{VSM}$. This deviation cannot arise from impurities in the sample, because an impurity would result in a reduced $M^{VSM}$ saturation magnetization. There is no apparent reason for the higher $M^{neutron}$ value in sample C. The other samples have $M^{neutron}$ and $M^{VSM}$ corresponding very well.

Average. To determine the reproducibility of the autoclave synthesis, an 'Average' column was added in Table 4. The 'Average' compares the statistical average of the refined values of all samples. As a statistical average is employed, the uncertainty of each parameter was based on the standard deviation method. Generally, the statistical average corroborates the individual parameters of each sample.

In summary, the comparison of the refinements of multiple data sets from the same sample shows hardly any variations, which clearly indicates that both the data collection of the DMC instrument and the Rietveld refinements yield strongly reproducible results. Additionally, by comparing the refined parameters between samples A, B and C we can conclude that the synthesis method is likewise reproducible, since the refined crystal/magnetic structures are almost identical with only minor deviations. The PXRD and NPD patterns do not show traces of impurities or eventual amorphous phase, and by comparing the theoretical magnetization ($M^{neutron}$) and the measured saturation magnetization ($M^{VSM}_{sat}$), we can argue that all samples are highly crystalline.

### 3.3. Part 3, effect of different synthesis approaches

For high-$Q$-coverage neutron data ($Q_{max} > 4$ Å), some additional degrees of freedom may be included in the refinement since more data are available. In order to investigate this statement, a study was performed to compare the effect of the ADP being described by one parameter ($B_{ov}$) or by two distinct isotropic displacement parameters ($B_{iso}$) for the metal ions and oxygen. This study is detailed in the supporting information (Tables S6 and S7) and was performed on part 3(a) data. In short, it was observed that the $B_{iso}$(Fe/Co) values were larger than the $B^{calc}_{ov}$ and caused a discrepancy between the Fe occupancy and $R_x$ between the $B_{iso}$ and $B_{ov}$ models. The $R_x$ in the $B_{iso}$ model is slightly higher than that in the $B_{ov}$, but within the uncertainties. The $B_{ov}$ model is reported in the following, as the $M^{neutron}$ values correspond better to the experimental $M^{VSM}$ data and the model is more similar to the model used in the previous parts of the paper. The $B_{iso}$ model, on the other hand, results in samples closer to the ideal stoichiometry of the CoFe$_2$O$_4$ spinel. The larger-$Q$-range data allowed the occupancy of the Td and Oh sites to be refined individually. This leads to the nominal formula (Co$^{2+}_{1-x}$Fe$^{3+}_x$)$^{Td}$[Co$^{2+}_y$Fe$^{3+}_{2-y}$]$^{Oh}$O$_4$, with $x$ and $y$ representing the inversion degrees for the Td and Oh sites, respectively. The formula unit charge balance was ignored for the obtained overall structure, when refining $x$ and $y$ freely.

(a) Effect of different hydrothermal reactors. This study is based on the work of Stingaciu et al. (2017), and the purpose is to investigate the crystal and magnetic structures of four cobalt ferrite spinel samples prepared using different hydrothermal reactors: AC, SR and FR. The main difference between these three reactors is the time it takes for the reactor to reach the desired reaction temperature (heating rate) and how long the sample is held at the elevated temperature. The slowest heating takes place in the AC ($\sim 0.15°C\,s^{-1}$), followed by the SR ($25°C\,s^{-1}$), while in the FR the heating is almost instantaneous ($\sim 500°C\,s^{-1}$). Four samples were investigated by PXRD and NPD using a Cu $K\alpha_{1,2}$ source Rigaku diffractometer and the neutron diffractometer PUS: two samples were made using the FR at 220°C (FR220) and 320°C (FR320), one from the SR at 240°C (SR240), and one from the AC at 240°C (AC240). The collected powder diffraction patterns of these samples are shown in Fig. 3. The presence of an impurity (main peaks at 1.7, 2.3 and 3.4 $Å^{-1}$) can be observed in Fig. 3(a), and it is identified as hematite, $\alpha$-$Fe_2O_3$ (space group $R\bar{3}c$, No. 167). The formation of $\alpha$-$Fe_2O_3$ cannot be avoided when using a precursor with pH below 12 (Cote et al., 2003; Zhao et al., 2007). However, regarding the quantity of $OH^-$ ions introduced by the sodium hydroxide solution, the pH is estimated as being equal to 15.2. Therefore the presence of hematite is unexpected, and the hematite content within sample FR220 is estimated to be <1 wt%.

The Cu source was used because the Co source had not yet been installed in the laboratory. The sloping background can be explained by the fluorescence signal decaying at a high angle ($Q > 5\,Å^{-1}$) and is clearly visible in all of the PXRD patterns in Fig. 3. From the results in part 1(b) and the refinements of model (iii) (DMC/Cu), it was expected that replacing DMC data with data from PUS, which has twice the $Q$ range (Table 2), would yield a robust description. Table 6 shows the refined parameters for the four samples; the refined parameters are discussed below.

Unit-cell parameter. The largest unit cell is obtained for samples AC240 (slowest heating rate), followed by SR240 (medium heating rate), while the two FR samples (fastest heating rate) exhibit smaller unit cells. In the literature (Cedeño-Mattei & Perales-Pérez, 2009; Andersen & Christensen, 2015; Cote et al., 2003), the unit-cell lengths for $CoFe_2O_4$ have previously been reported to be between 8.31 and 8.43 Å, and all the values found here are within this range.

On the basis of the *in situ* studies by Andersen & Christensen (2015) on the formation of $CoFe_2O_4$, the largest unit-cell parameters were expected for the smallest crystallites. However, the opposite trend is seen here, suggesting the unit-cell dimension is not only related to the crystallite size but also very dependent on the stoichiometry and specific site occupation of Co/Fe and potential vacancies.

Apparent crystallite size. Stingaciu et al. (2017) reported that heating rates, pressure and temperature play an important

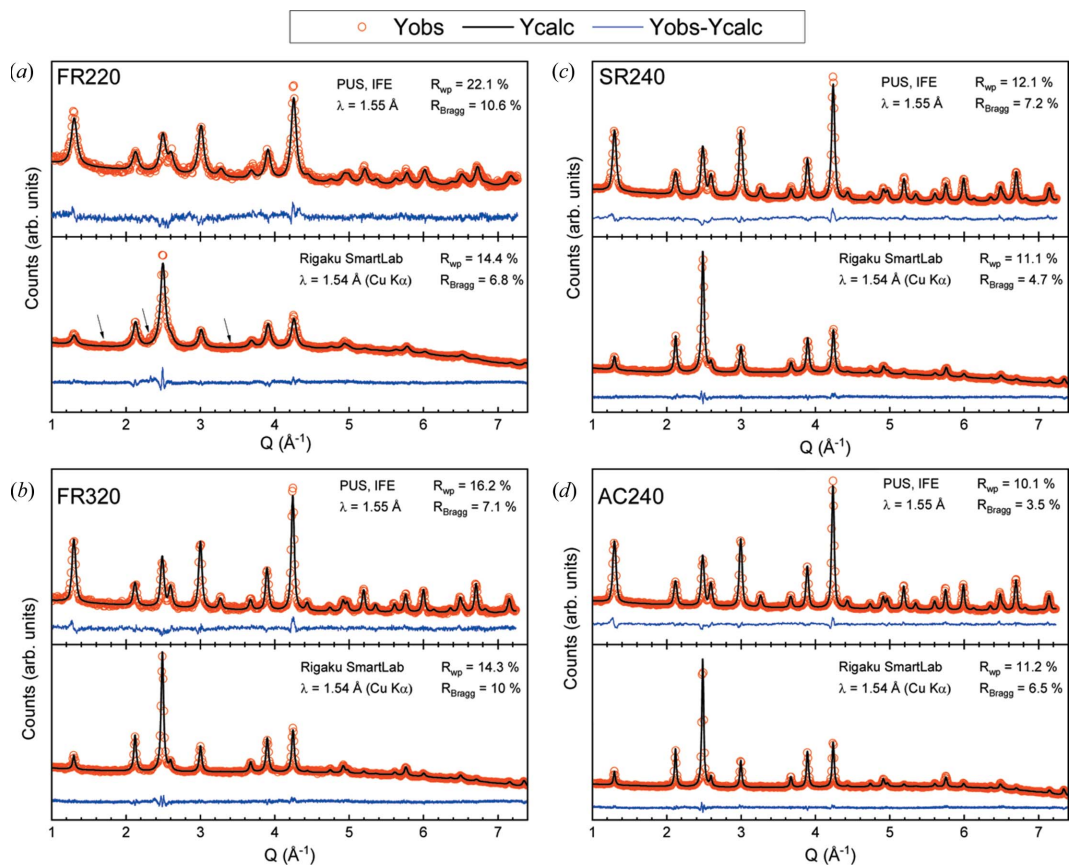

**Figure 3**
(a) FR220, (b) FR320, (c) SR240 and (d) AC240. The data are shown by the red circles, the refined model by the black line and the residual by the blue line. Weighted profile and Bragg factors, respectively $R_{wp}$ and $R_{Bragg}$, are indicated for each diffraction pattern. Black arrows show the three main contributions of hematite, $\alpha$-$Fe_2O_3$. Only frequencies of 3 and 15 data points were selected to draw the NPD and PXRD patterns, respectively.

**Table 6**
Refinement comparison of four samples synthesized by either FR, SR or AC.

The temperature used during the synthesis is included in the name of the sample. The refinement model employed a combination of PXRD data (Cu $K\alpha$) and NPD data from PUS. Due to the high $Q$ range of the NPD data, the Td and Oh site occupancies could be refined independently. The distribution of both $Fe^{3+}$ and $Co^{2+}$ within Td and Oh sites is detailed as $x$ and $y$, respectively. $m_{f.u.}$ and $M^{neutron}$ were calculated on the basis of the refined site occupancy and $R_x$. The X-ray and magnetic data were previously published by Stingaciu *et al.* (2017), but the neutron data and all refinements have not been previously published.

| | FR220 | FR320 | SR240 | AC240 |
|---|---|---|---|---|
| Unit cell (Å) | 8.3532 (4) | 8.3785 (2) | 8.3866 (2) | 8.3925 (1) |
| Crystallite size (nm) | 5.2 (8) | 10.9 (8) | 10.7 (8) | 15.1 (8) |
| Crystallite size (nm) (Stingaciu *et al.*, 2017) | 8.2 (1) | 10.6 (1) | 11.6 (1) | 15.3 (1) |
| $x$ (O) | 0.2417 (2) | 0.2413 (1) | 0.2419 (1) | 0.2423 (1) |
| $B_{ov}$ (Å$^2$) | 1.40 (2) | 1.20 (2) | 1.14 (2) | 1.00 (1) |
| Occ($Co^{2+}$)$^{Td}$ (%) | 35 (2) | 37 (3) | 38 (1) | 33 (1) |
| Occ($Fe^{3+}$)$^{Td}$ (%) | 65 (3) | 63 (4) | 62 (1) | 67 (1) |
| Occ($Co^{2+}$)$^{Oh}$ (%) | 45 (2) | 39 (1) | 41 (1) | 42 (1) |
| Occ($Fe^{3+}$)$^{Oh}$ (%) | 55 (3) | 61 (1) | 59 (1) | 58 (1) |
| ($Co^{2+}_{1-x}Fe^{3+}_x$)$^{Td}$ | (Co$_{0.35 (2)}$Fe$_{0.65 (3)}$) | (Co$_{0.37 (3)}$Fe$_{0.63 (4)}$) | (Co$_{0.38 (1)}$Fe$_{0.62 (2)}$) | (Co$_{0.33 (1)}$Fe$_{0.67 (1)}$) |
| [$Co^{2+}_y Fe^{3+}_{2-y}$]$^{Oh}$ | [Co$_{0.90 (2)}$Fe$_{1.10 (3)}$] | [Co$_{0.78 (1)}$Fe$_{1.22 (2)}$] | [Co$_{0.81 (1)}$Fe$_{1.19 (1)}$] | [Co$_{0.84 (1)}$Fe$_{1.16 (2)}$] |
| Co:Fe ratio | 1.26 (5):1.74 (6) | 1.16 (3):1.84 (5) | 1.19 (1):1.81 (2) | 1.17 (1):1.83 (2) |
| $R_x(Co^{2+})^{Oh}$ ($\mu_B$) | 2.10 (3) | 2.16 (2) | 2.15 (1) | 2.22 (1) |
| $R_x(Fe^{3+})^{Oh}$ ($\mu_B$) | 3.49 (6) | 3.59 (4) | 3.58 (2) | 3.70 (2) |
| $m_{f.u.}$ ($\mu_B$ f.u.$^{-1}$) | 2.7 (2) | 3.0 (2) | 3.0 (1) | 3.0 (1) |
| $M^{neutron}$ (A m$^2$ kg$^{-1}$) | 65 (6) | 72 (4) | 70 (2) | 70 (2) |
| $M^{VSM}_{sat}$ (A m$^2$ kg$^{-1}$) | 38.68 (2) | 66.33 (2) | 62.77 (2) | 68.58 (2) |
| $R_{wp}$ (%) | 22.6/14.5 | 16.5/14.5 | 12.4/11.5 | 10.3/11.2 |
| $\chi^2$ | 1.6/2.5 | 1.3/1.8 | 3.7/1.5 | 2.1/1.5 |
| $R_{Bragg}$ (%) | 11.0/6.6 | 7.5/10.3 | 7.5/4.5 | 3.8/6.9 |
| $R_{mag}$ (%) | 13.6/– | 9.26/– | 8.84/– | 4.56/– |
| #reflections | 24/24 | 24/24 | 24/24 | 24/24 |

role in controlling the size of the nanoparticles. The heating rate and holding time effects were seen for the unit-cell parameter and the crystallite size. AC240 has the largest crystallite size among the four samples, while SR240 and FR320 have identical ACS. Also, the obtained values are in agreement with the data from Stingaciu *et al.* (2017). The study can affirm that increasing the temperature drastically increases the ACS, *e.g.* FR320 is twice as large [10.9 (8) nm] as FR220 [5.2 (8) nm].

Overall isotropic displacement parameter ($B_{ov}$). All samples have a different $B_{ov}$ value, with FR220 having the highest, FR320 and SR240 being similar, and AC240 having the lowest. Whether the observed decrease throughout the four samples is an effect of the crystallite size is difficult to conclude from the available data. In fact, this effect could be explained by the fact that the ADPs are strongly correlated to background refinement. Nevertheless, the obtained values are within the range expected for inorganic compounds (~0.5 to ~3 Å$^2$) (Pecharsky & Zavalij, 2009).

Site occupancy. All samples have similar site occupancy, with $Fe^{3+}$ occupying almost 2/3 of Td sites, which is close to a random occupancy for a stoichiometry of Fe:Co = 2:1 (Sickafus *et al.*, 1999; Sorescu *et al.*, 2021) and 60% of Oh sites. Only FR220 deviates slightly, but it is still close to other samples when considering the uncertainty.

Magnetic properties. Even though the four samples have similar atomic magnetic moments, a trend emerges where FR320 and SR240 have equal $R_x$, while AC240 has the highest value and FR220 the lowest. This trend could be due to the increasing ACS. Indeed, it is found in the literature that increasing the particle diameter enhances the saturation magnetization (Stingaciu *et al.*, 2017; Andersen *et al.*, 2019; López-Ortega, Lottini *et al.*, 2015; Maaz *et al.*, 2007). Despite variations in $R_x$, FR320, SR240 and AC240 have the same $m_{f.u.}$ (3.0 $\mu_B$ f.u.$^{-1}$), while for FR220 it is lower (2.7 $\mu_B$ f.u.$^{-1}$), but within the standard deviation. The differences are explained by variations in the occupancy of the different sites and the Co/Fe ratio. Comparing FR220 and FR320, it is revealed that increasing the temperature enhances the magnetic properties.

Regarding the saturation magnetization obtained from VSM measurements, a large discrepancy is seen in the FR220 and FR320 samples: $M^{VSM}_{sat}$(FR320) is almost twice as large as $M^{VSM}_{sat}$(FR220). Concerning the FR220 sample, the reason why $M^{VSM}_{sat}$ is significantly smaller than the others (38.68 A m$^2$ kg$^{-1}$) is probably the small ACS, which increases the surface area to volume ratio, thus increasing the relative amount of surface structural reordering. It could also be due to cation vacancies and defects in the octahedral sites, reducing the magnetic moment (Huang *et al.*, 2017). With the present structural model, vacancies on the octahedral site would be modelled as an increased $Co^{2+}$ occupancy on the octahedral site. Another explanation for the low saturation magnetization of FR220 is that the sample is partly superparamagnetic, since the ACS is small enough to allow such magnetic behaviour (Alzoubi *et al.*, 2020; Mooney *et al.*, 2004; Ahn *et al.*, 2003; Moumen & Pileni, 1996; Sangeneni *et al.*, 2018). An impurity phase of $\alpha$-Fe$_2$O$_3$ was also identified in FR220, although the content was <1 wt%. Hematite is an antiferromagnet/weak ferromagnet at ambient temperature (Tadic *et al.*, 2014; Aharoni *et al.*, 1962). However, the possibility cannot be excluded that other non-crystalline phases are contributing to the mass of the sample. Such impurities can explain the low $M^{VSM}_{sat}$ observed for FR220. It is also noted that the remaining samples have higher $M^{neutron}$ than $M^{VSM}_{sat}$. Only the values for AC240 are close, with a deviation of just 2 A m$^2$ kg$^{-1}$, compared with 6 and 8 A m$^2$ kg$^{-1}$ for FR320 and SR240, respectively.

In parts 1(*b*) and 2 it was found that the AC synthesis produced highly crystalline materials, as corroborated by the values of $M^{neutron}$ and $M^{VSM}_{sat}$. The most likely explanations for the discrepancies for the other samples are the presence of non-crystalline impurities adding to the mass of the sample. Additionally, for FR220 the model may not correctly describe the system, because the interstitial sites are forced to be fully occupied and vacancies could occur on the octahedral site.

Table 7
List of refined parameters for CoFe$_2$O$_4$ samples prepared with different Co-containing salts.

The refinement model uses a combination of in-house PXRD data (Cu $K\alpha$) and NPD data from HRPT. Occupancies of the Td and Oh sites were refined separately. The saturation magnetization extracted from VSM measurements ($M_{sat}^{VSM}$) is tabulated along with the calculated macroscopic magnetization ($M^{neutron}$). None of this work has been previously published.

| | Co(Ac)$_2$ | Co(NO$_3$)$_2$ | CoCl$_2$ |
|---|---|---|---|
| Unit cell (Å) | 8.4031 (2) | 8.4058 (2) | 8.4060 (2) |
| Crystallite size (nm) | 10.2 (8) | 10.6 (8) | 10.7 (8) |
| $x$ (O) | 0.2430 (1) | 0.2430 (1) | 0.2430 (1) |
| $B_{ov}$ (Å$^2$) | 0.65 (1) | 0.74 (1) | 0.69 (1) |
| Occ(Co$^{2+}$)$^{Td}$ (%) | 39 (1) | 40 (1) | 39 (1) |
| Occ(Fe$^{3+}$)$^{Td}$ (%) | 61 (1) | 60 (1) | 61 (1) |
| Occ(Co$^{2+}$)$^{Oh}$ (%) | 39 (1) | 40 (1) | 40 (1) |
| Occ(Fe$^{3+}$)$^{Oh}$ (%) | 61 (1) | 60 (1) | 60 (1) |
| (Co$^{2+}_{1-x}$Fe$^{3+}_x$)$^{Td}$ | (Co$_{0.39(1)}$Fe$_{0.61(1)}$) | (Co$_{0.40(1)}$Fe$_{0.60(1)}$) | (Co$_{0.39(1)}$Fe$_{0.61(1)}$) |
| [Co$^{2+}_y$Fe$^{3+}_{2-y}$]$^{Oh}$ | [Co$_{0.78(1)}$Fe$_{1.22(2)}$] | [Co$_{0.80(1)}$Fe$_{1.20(2)}$] | [Co$_{0.80(1)}$Fe$_{1.20(2)}$] |
| Co:Fe ratio | 1.17 (1):1.83 (2) | 1.20 (1):1.80 (2) | 1.19 (1):1.81 (2) |
| $R_x$(Co$^{2+}$)$^{Oh}$ ($\mu_B$) | 2.32 (1) | 2.26 (1) | 2.29 (1) |
| $R_x$(Fe$^{3+}$)$^{Oh}$ ($\mu_B$) | 3.87 (2) | 3.77 (2) | 3.81 (2) |
| $m$ ($\mu_B$ f.u.$^{-1}$) | 3.3 (1) | 3.2 (1) | 3.2 (1) |
| $M^{neutron}$ (A m$^2$ kg$^{-1}$) | 77 (2) | 75 (2) | 76 (2) |
| $M_{sat}^{VSM}$ (A m$^2$ kg$^{-1}$) | 66.81 | 69.32 | 66.67 |
| $R_{wp}$ (%) | 8.6/9.8 | 9.1/9.5 | 8.9/9.3 |
| $\chi^2$ | 2.9/1.3 | 3.1/1.4 | 3.0/1.3 |
| $R_{Bragg}$ (%) | 2.9/4.3 | 3.5/2.4 | 3.4/3.3 |
| $R_{mag}$ (%) | 4.9/– | 5.6/– | 5.5/– |
| #reflections | 31/21 | 31/21 | 31/21 |

In summary, the heating rate, holding time and temperature play an important role in controlling the size of the crystallites, and generally in the atomic and microstructural properties of a sample. However, these reaction parameters appear to be less influential when it comes to the site occupancy of Co/Fe. The sample standing out in this study is FR220, which has a reduced size and a smaller saturation magnetization compared with the other samples. Samples FR320, SR240 and AC240 are relatively similar: only the unit cell and crystallite size differ. AC240 exhibits the largest crystallite size and the highest saturation magnetization.

(*b*) Different cobalt salts. In addition to investigating the influence of the different hydrothermal reactors, the effect of using different cobalt salt precursors has also been studied as a potential way of influencing the Co/Fe site occupancy. Three different precursors were prepared for an AC reactor from three different cobalt salts (see Section 2), cobalt(II) acetate [Co(Ac)$_2$], cobalt(II) nitrate [Co(NO$_3$)$_2$] and cobalt(II) chloride (CoCl$_2$). The samples were characterized by PXRD using Cu $K\alpha_{1,2}$ radiation and NPD at HRPT at SINQ. The diffraction patterns can be found in the supporting information (Fig. S6), while the parameters extracted from the refinements are shown in Table 7. The effect of $B_{ov}$ versus $B_{iso}$ was also investigated in this study, and only minor differences were found between the two models, where a lowering of $M^{neutron}$ by 1 A m$^2$ kg$^{-1}$ is the most striking difference.

Following the results shown in Table 7, no significant deviations are observed between the samples using cobalt acetate, cobalt nitrate or cobalt chloride as salt precursor. It can be concluded that the different salts have a minor effect on the resulting crystal structure and crystallite size, *i.e.* the counter-ion NO$_3^-$, Cl$^-$ or CH$_3$COO$^-$ plays an insignificant role in the hydrothermal synthesis of CoFe$_2$O$_4$. Note that all samples have the same stoichiometry and site occupation of Co/Fe on the various sites in the structure (40/60% in all sites). The refined magnetic moments are comparable for all samples, but the calculated $M^{neutron}$ significantly exceeds the value obtained from VSM measurements. The $M^{neutron}$ values are identical to those calculated using model (iii) (DMC/Cu) in part 1(*b*). HRPT provides significantly more information compared with DMC due to the wider $Q$ range, but the magnetic information is primarily found at low $Q$, also covered by DMC. The discrepancy between $M^{neutron}$ and $M_{sat}^{VSM}$ is probably explained by a non-magnetic amorphous phase in the sample. In agreement with what was observed in parts 1(*b*), 2 and 3(*a*) for the samples synthesized with the AC reactor, none of the diffraction patterns (Fig. S6) shows traces of crystalline impurities.

In summary, the crystal and magnetic structures of CoFe$_2$O$_4$ nanoparticles prepared from three different precursors [containing Co(OAc)$_2$, CoCl$_2$ and Co(NO$_3$)$_2$] have been compared. The three samples exhibit similar ACS, oxygen positions and site occupancies. Only the unit-cell parameter, thermal vibrations and magnetic moments reveal slight differences. The three samples are also magnetically similar and their net intrinsic magnetization is almost identical, as is their macroscopic magnetization ($M_{sat}^{VSM}$), further evidence that the investigated precursor anions are not influencing the final product in the autoclave synthesis. The observed difference between $M^{neutron}$ and $M_{sat}^{VSM}$ may be due to the presence of amorphous phases or size and/or size distribution effects with regards to macroscopic magnetic properties.

### 3.4. Part 4, effect of synthesis conditions

This part of the paper features the four CoFe$_2$O$_4$ samples synthesized by the autoclave reactor, which were prepared from Co(NO$_3$)$_2$·6H$_2$O and Fe(NO$_3$)$_3$·9H$_2$O in a ratio of 1:2. The four samples were characterized using three different NPD instruments and two different in-house X-ray diffractometers.

Table 8 gives a summary of the obtained structural parameters. The first two columns are extracted from part 1 and show model (ii) (DMC/Co) and model (iii) (DMC/Cu), while the third column is the average column from Table 5. The results are in general remarkably similar, with minor deviations related to crystallite size and unit cell, which can be attributed to the differences in synthesis conditions. The refined occupancies on the octahedral sites are nearly identical, while larger deviations are found for the tetrahedral site,

**Table 8**
Some refined structural parameters of selected samples from previous tables, including unit cell, crystallite size, oxygen coordinates, occupancy of the Td and Oh sites, and the refined magnetic moment along $R_x$.

All samples summarized here were synthesized using an autoclave reactor.

| Sample | Model (ii) DMC/Co | Model (iii) DMC/Cu | Average of A, B and C | AC240 | Co(NO$_3$)$_2$ |
|---|---|---|---|---|---|
| Temperature/time (°C/h) | 240/2 | 240/2 | 240/2 | 240/1 | 200/1 |
| Fe:Co:NaOH (M) | 2:2:16 | 2:2:16 | 3:2.3:16 | 2:2:16 | 2:1:12 |
| OH$^-$:NO$_3^-$ | 2:1 | 2:1 | 1.25:1 | 2:1 | 2.25:1 |
| Pattern | DMC/Co | DMC/Cu | DMC/Co | PUS/Cu | HRPT/Cu |
| Unit cell (Å) | 8.3890 (1) | 8.3891 (3) | 8.3919 (6) | 8.3925 (1) | 8.4058 (2) |
| Crystallite size (nm) | 13.3 (8) | 13.1 (8) | 13.0 (3) | 15.1 (8) | 10.6 (8) |
| $x$ (O) | 0.2428 (1) | 0.2411 (3) | 0.2434 (2) | 0.2423 (1) | 0.2430 (1) |
| $B_{ov}$ (Å$^2$) | 1.22 (1) | 0.89 (4) | 1.50 (6) | 1.00 (1) | 0.74 (1) |
| Occ(Co$^{2+}$)$^{Td}$ (%) | 24 (1) | 29 (1) | 19 (1) | 33 (1) | 40 (1) |
| Occ(Fe$^{3+}$)$^{Td}$ (%) | 76 (1) | 71 (1) | 81 (1) | 67 (1) | 60 (1) |
| Occ(Co$^{2+}$)$^{Oh}$ (%) | 38 (1) | 35 (1) | 40 (1) | 42 (1) | 40 (1) |
| Occ(Fe$^{3+}$)$^{Oh}$ (%) | 62 (1) | 65 (1) | 60 (1) | 58 (1) | 60 (1) |
| Co:Fe ratio | 1.00 (1):2.00 (2) | 1.00 (1):2.00 (2) | 1.00 (2):2.00 (2) | 1.17 (1):1.83 (2) | 1.20 (1):1.80 (2) |
| $R_x$(Co$^{2+}$)$^{Oh}$ ($\mu_B$) | 2.33 (1) | 2.31 (1) | 2.5 (1) | 2.22 (1) | 2.26 (1) |
| $R_x$(Fe$^{3+}$)$^{Oh}$ ($\mu_B$) | 3.88 (2) | 3.86 (2) | 4.2 (2) | 3.70 (2) | 3.77 (2) |
| $m_{f.u.}$ ($\mu_B$ f.u.$^{-1}$) | 3.1 (1) | 3.2 (1) | 3.2 (2) | 3.0 (1) | 3.2 (1) |
| $M^{neutron}$ (A m$^2$ kg$^{-1}$) | 73 (3) | 77 (2) | 75 (4) | 70 (2) | 75 (2) |
| $M_{sat}^{VSM}$ (A m$^2$ kg$^{-1}$) | 73.5 (2) | 73.5 (2) | 73.5 (6) | 68.58 (2) | 69.32 |

especially for the average of samples A, B and C (third column), as well as the sample synthesized at 200°C (last column). Some of the differences may also be attributed to the Co:Fe ratio being fixed in the refinement of the DMC data, which does not allow deviations from the nominal composition or introduction of vacancies. The thermal vibration also deviates, and since this parameter correlates strongly with the occupancies it may also help explain the discrepancy. Regarding the atomic magnetic dipole moment, the average of samples A, B and C from part 2 is the sample that exhibits the highest $R_x$ values compared with the three other samples, which have similar values. For the calculated net intrinsic magnetization, which is a parameter depending on the refinement method used, we clearly see that the samples have close to identical values [73 (3)–77 (2) A m$^2$ kg$^{-1}$]. Concerning the saturation magnetization, two sets of values are observed: AC240 and Co(NO$_3$)$_2$ exhibit close to identical values, and the samples from parts 1(b) and 2 are identical. The AC240 and Co(NO$_3$)$_2$ samples were held for only 1 h, while all other samples were reacted for 2 h. The reduced reaction time is likely to cause these samples to have lower crystallinity due to the introduction of disorder, vacancies or amorphous impurity phases. It has been reported that performing the autoclave reaction for 19 h increases the saturation magnetization while reducing the reaction temperature to 100°C reduced the saturation magnetization (Stingaciu et al., 2017). The neutron diffraction data reveal that the crystalline part of the nanoparticles is largely unaffected by the synthesis conditions.

## 4. Conclusion

The present study consists of four distinct parts, each dedicated to examining the robustness, reproducibility and reliability of structural parameters obtained from Rietveld analysis of powder diffraction data (X-ray and neutron). Nanocrystalline cobalt ferrite, CoFe$_2$O$_4$, was used as a sample prepared using different synthesis conditions and collected using different instrumentation and radiation sources. The study devotes special attention to the reliability of the site occupancies of Co$^{2+}$ and Fe$^{3+}$ in CoFe$_2$O$_4$, as the atomic structure is the key parameter for examining the structure–property relationship in spinel ferrites. In part 1(b) it was demonstrated that combining X-ray powder diffraction from Cu and Co sources with low-$Q$-range neutron powder diffraction data gives reliable Fe/Co occupancies and magnetic structure for CoFe$_2$O$_4$. Reliable Fe/Co occupancies were obtained from refining Co-source data solely. In other words, the in-house powder pattern has sufficient contrast to distinguish the neighbouring elements of Fe and Co, when taking into consideration the resonant scattering contributions $f'$ and $f''$. This may be expanded to other elements, e.g. Fe/Mn and Fe/Ni, allowing in-house determination of site occupancies in spinel structures and other transition metal oxides with mixed occupancies, e.g. battery materials. Here neutron diffraction data improved the robustness of the model, even if only six Bragg peaks were included.

Part 2 was focused on the reproducibility of the synthesis, data collection and refinements of the low-$Q$-range neutron data. Comparing multiple measurements on the same sample clearly showed that the low-$Q$-range neutron data collection and Rietveld refinements are highly reproducible. The comparison of three samples with identical synthesis procedure showed that the synthesis method was reproducible, with only minute deviations.

Part 3(a) compared three samples prepared using different hydrothermal reactors, with different heating rates, namely, a continuous flow reactor (FR), spiral reactor (SR) and autoclave (AC). The size is the parameter most affected, while the distribution of Co/Fe between the octahedral and tetrahedral sites in the structure is less influenced by the heating rate and holding time.

Part 3(b) was dedicated to the investigation of the influence of using different cobalt salts in the precursors: $CoCl_2$, $Co(NO_3)_2$ and $Co(CH_3COOH)$. The three $CoFe_2O_4$ nanoparticle samples prepared from the different precursors have practically identical crystal and magnetic structures, demonstrating that the cobalt salt anions used have no influence on the final product.

Finally, part 4 described considerations regarding differently prepared samples of $CoFe_2O_4$ nanoparticles using an autoclave. Despite differences in the macroscopic magnetic properties, all samples exhibited similar crystal and magnetic structure.

The presented study revealed that the hydrothermal synthesis of $CoFe_2O_4$ is highly reproducible. The net crystallographic magnetizations calculated from refined occupancies and the atomic dipolar magnetic moment are generally in good agreement with macroscopic magnetic measurements, except for the smaller crystallite where superparamagnetic behaviour or disorder and amorphous phases may play a role. Reliably establishing this link between crystal/magnetic structure and observed magnetic properties is key to investigating the structure–property relationship in spinel ferrites as well as other magnetic compounds.

## 5. Related literature

The following references are cited in the supporting literature: Langford & Wilson (1978), Rodríguez-Carvajal (2003), Shanmugavani *et al.* (2015).

### Acknowledgements

We gratefully acknowledge neutron beamtime at PSI and IFE and support by the beamline staff Emmanuel Canévet and Denis Sheptyakov at PSI and Magnus Sørby at IFE. Affiliation with the ESS lighthouses SMART and Q-MAT as well as the Center for Integrated Materials Research (iMAT) at Aarhus University is gratefully acknowledged.

### Funding information

MC and HLA are grateful for support from the Carlsberg Foundation (grant Nos. CF16-0084, CF18-0519 and CF19-0143). CG-M acknowledges financial support from MICINN through the 'Juan de la Cierva' Program (FJC2018-035532-I). JVA acknowledges financial support from Nordforsk (project No. 106874). We thank the Danish Agency for Science, Technology and Innovation for funding the instrument center DanScatt (7129-00003B).